\documentclass{icrc}

\usepackage{times}
 \usepackage{graphicx} 

\begin{document}

\title{Very High Energy $\gamma-$ ray Emission from Crab Nebula with the PACT Array}
\author[1]{P.R.Vishwanath}
\author[1]{B.S.Acharya}
\author[1]{P.N.Bhat}
\author[1]{V.R.Chitnis}
\author[1]{P.Majumdar}
\author[1]{M.A.Rahman}
\author[1]{B.B.Singh}
\affil[1]{Tata Institute of Fundamental Research,Colaba\\
Mumbai,400005, India}

\correspondence{vishwa@tifr.res.in}

\runninghead{Vishwanath et al.: $\gamma-$ ray emission from Crab Nebula with PACT}
\firstpage{1}
\pubyear{2001}


\maketitle

\begin{abstract}
The Crab nebula has proved to be the nearest to a standard candle in VHE $\gamma-$ ray
astronomy. Results on the gamma ray emission from the nebula at various energies  have
come in the last decade mostly from imaging telescopes. The aim of the new Pachmarhi
Atmospheric Cerenkov Telescope (PACT) array has been to use the temporal and spatial
distribution of Cerenkov photons in distinguishing between proton and gamma ray
showers.  The array, with timing information from 175 mirrors, is ideally suited for
precise estimation of the arrival direction. Preliminary results from the  recent data taken
on Crab  has shown a 12$\sigma$ signal with the flux in good agreement with those of the
other experiments.
\end{abstract}

\section{Introduction}
Crab nebula has been the most intensively studied object in all High Energy
Astrophysics. There were many attempts to detect VHE gamma ray emission from the
Crab in the 70s and early to mid 80s with varying success. Its certain detection in the
late '80s by the Whipple Imaging telescope [\cite{Wee88}] established the field of VHE
$\gamma-$ ray astronomy on a firm basis. Since then many groups including Whipple have
been able to establish good signals from the source. The energy range from 100 GeV
to 7000 GeV has been covered by the Atmospheric Cerenkov telescopes. There have also
been detection of Crab in this energy range by EAS experiments. Further, the VHE
$\gamma-$ rays from Crab have been explained as inverse Compton scattering of 
relativistic
electrons off soft neighbouring photons. [\cite{Ong98}]

\section{PACT Experiment}
A new atmospheric Cerenkov array to study cosmic sources of Very High Energy
(VHE) $\gamma$ rays has been set up in Pachmarhi in central India. While most of the
atmospheric Cerenkov experiments in the recent years are based on the imaging technique,
the aim of the new Pachmarhi Atmospheric Cerenkov Telescope (PACT) array has been
to use the temporal and spatial distribution of Cerenkov photons in distinguishing
between proton and gamma ray showers for increase of sensitivity. The array  consists of
25 telescopes deployed in a field of 80m $\times$ 100 m area [\cite{Bha00}]. Each telescope
consists of 7 parabolic mirrors, each of diameter 0.9 m. Each mirror is looked at by a
fast PMT behind a 3 degree circular mask. To minimize the attenuation and distortions in
the cables, the array is divided into 4 sectors, with each sector servicing a group of 6
nearby telescopes. The individual mirror rates were kept at about 5 KHz. The pulses
from the 7 PMTs in a telescope were added linearly to form a `Royal Sum' pulse for each
telescope. These Royal Sum pulses were discriminated to give a rate of about 30 KHz. A
four fold coincidence of these Royal Sum pulses generated the event trigger. A single
sector event rate was about 3 to 4 Hz for most atmospheric conditions. Both the timing
and pulse height information for each mirror are  recorded in the nearby sector
electronics hut serving each sector. All the royal sum pulses were brought into the control
room and their arrival times recorded along with the other housekeeping
information. With all 4 sectors functioning, the overall event rate was about 9 Hz. Monte
Carlo estimates of the energy threshold for one sector is 900 GeV. When at least 2
sectors are demanded in coincidence, the threshold is slightly higher. A total of 60 hours
of data was taken on the Crab nebula and about 20 hours on  several background regions,
but with the same declination as Crab.  The Crab data ranged from runs in Nov 1999 to
Nov 2000.





\section{Estimation of Arrival Direction}
The arrival direction has been calculated using the TDC information from the so
called Royal Sum pulses and individual mirror outputs. Several different methods have
been used for the estimation of the arrival direction [\cite{Maj01}]. We discuss
here the preliminary results from one method and the results from other methods will be
presented in the conference. The data used is from the more recent runs in fall of 2000
and covers 31 hours (15 nights) and 16 hours (10 nights) on source and background
respectively when two southern sectors were completely operational. Further it was
required that the events be common in both sectors to increase the number of degrees of
freedom and get a longer lever arm. The usable event rate was 1.3 Hz.
\begin{figure}[t]
\includegraphics[width=8.3cm]{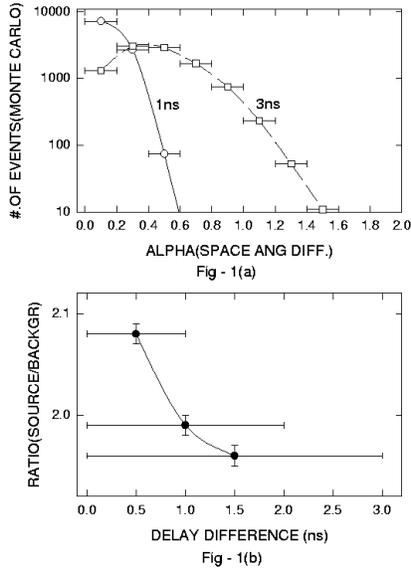} 
\caption{(a) $\alpha$ distributions for Monte Carlo generated events assuming HWHM of 1(left)
and 3(right)ns.
(b) The ratio of number of source to background events as function of the differences in
expected and observed delays in telescopes.}
\end{figure}
 A simple Monte Carlo program was written to understand the angular
distribution capabilities of the array. The arrival times  at all the 12 telescopes (in the
southern half of the array) were generated (for a point source at different zenith angles)
assuming a TDC distribution with different values for HWHM. Fig 1(a) shows the result of
angular reconstruction of Monte Carlo events assuming HWHM of 1 and 3 ns.
For a TDC distribution of HWHM equal to 1(3) ns, it was found that  73\% of the
events were within 0.25(0.6)$^\circ$  from the assumed source direction. Thus, $\Delta$$\tau$, the
absolute value of the difference between the expected and the observed arrival time of the
Cerenkov photons at the telescope is an indication of the offset of the direction of the
event wrt the source direction. Before the actual angular reconstruction process, all
events were binned as to the number of telescopes with $\Delta$$\tau$ $<$ 1, 2 and 3 ns. For
eg; there were 146492(74904) events in the data from source(background) with at least 4 telescopes
showing $\Delta$$\tau$ $<$ 3 ns. The Ratio R, of Source to background events where $\Delta$$\tau$, the
delay difference in at least 4 telescopes was $<$ 3 ns is 1.96$\pm$0.1. Fig 1(b) shows
this ratio R vs the delay difference. It is clearly seen that there is a considerable excess of
source events when $\Delta$$\tau$ is $<$ 1 ns. The excess, assuming the normalization from the
data at $<$ 3 ns, amounts to 4244$\pm$381events.
\par The standard angular reconstruction program was used to find $\alpha$, the space
angle difference between the events and the source direction. Fig 2(a) shows the $\alpha$
distribution for both on Crab and off Crab (normalized to total number of events on source)
directions. Except for differences at small $\alpha$, the distributions are quite similar. It
should be noted that these distributions are broader than those of Monte Carlo shown in
Fig 1(a). While the Monte Carlo calculations needed all the 12 telescopes, the data analysis
demanded at least 4 telescopes including  at least one at either extreme of the array to
have $\Delta$$\tau$ $<$ 1 ns. Fig 2(b) shows the excess events (source - background) as a
function of $\alpha$. The total excess within 0.6$^\circ$ amounts to 4428 $\pm$ 353 events, a
12$\sigma$ effect. The minutely rate is 2.4 $\pm$  0.2. When trigger from at least 2 sectors
are demanded, the energy threshold increases to 1100 GeV. The excess from the Crab
direction translates to a flux of $3.2 \pm 0.3 \times  10^{-11} phsec^{-1}cm^{-2}$
at $>$1100 GeV.

\begin{figure}[t]
 \vspace*{2.0mm} 
\includegraphics[width=8.3cm]{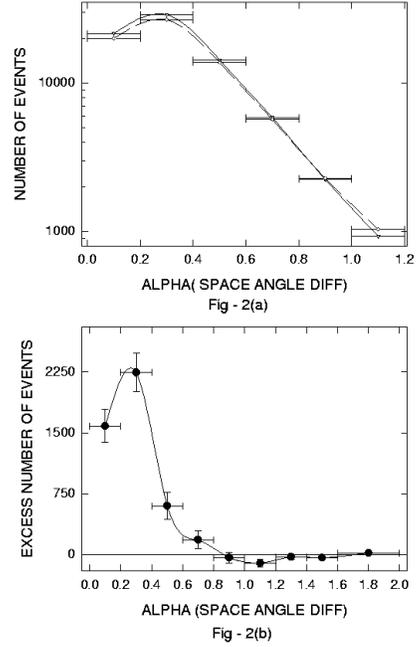} 
\caption{(a) $\alpha$ distributions for events in data source(upper) and background (lower) 
(b) The excess number of events v/s $\alpha$.}
\end{figure}

\section{Discussions}
The VHE $\gamma-$ ray flux at $>$1 TeV is $2.1 \pm 0.2 \pm 0.3$ $\times$ 
$10^{-11}$\\ 
$sec^{-1}cm^{-2}$
according to the fit given by the Whipple experiment. [\cite{Hil98}] Given
the preliminary nature of the above results from the Pachmarhi array, the derived flux for
the Crab is quite encouraging. However, one should note that the PACT array is capable
of better angular resolution and the methods being developed should increase the
sensitivity and get more precise value for the flux. Other methods, for eg, using the
timing jitter [\cite{Chi99}],[\cite{Chi01}], lateral distribution parameters etc, are 
also being evolved. Single sector
events would help go down in energy threshold. The trigger rate with the whole array (
with all the 4 sectors operating) is about 50\% more than with just 2 sectors. Thus the
data to be taken in the coming seasons should be able to see a strong signal from Crab in
a much shorter time.





%
\begin{figure}[t]
\end{figure}

%




\begin{acknowledgements}
It is a pleasure to thank Sarvashri A.I.DSouza, J.Francis, K.S.Gothe, 
B.K.Nagesh,
M.S.Pose, P.N.Purohit, K.K.Rao, S.K.Rao, S.K.Sharma, A.J.Stanislaus,
P.V.Sudershanan, S.S.Upadhyay, B.L.Venkatesh Murthy for their participation 
in various
aspects of the experiment.

\end{acknowledgements}




\end{document}